\documentstyle[epsf]{l-aa}
\newcommand{\be}{\begin{equation}}
\newcommand{\ee}{\end{equation}}

\begin{document}

\thesaurus{06   
        (13.07.3; 13.07.2; 08.05.1; 08.23.3)}
\title{Can the gamma-ray source 3EG J2033+4118 be produced by the stellar
system Cyg OB2 No. 5?}
\author{P. Benaglia\inst{1}\thanks{Member
of CONICET}, G.E. Romero\inst{1,\star}, I.R. Stevens\inst{2}, and
D.F. Torres\inst{1,}\thanks{Fellow of CONICET}} \offprints{P.
Benaglia}

\institute{Instituto
 Argentino de Radioastronom\'{\i}a, C.C.5,
(1894) Villa Elisa, Bs.\ As., Argentina \and School of Physics \&
Astronomy, University of Birmingham, Edgbaston, Birmingham B15
2TT, UK}

\date{\today}
\maketitle

\markboth{P.Benaglia et al.: Gamma-rays from Cyg OB2 No. 5}{}

\begin{abstract}
We discuss the possibility that the stellar system Cyg OB2 No. 5
can be a gamma-ray source in the light of recent EGRET and radio
data. This system is formed by an O7 Ia + Ofpe/WN9 contact binary.
A third star, probably a B0 V star also associated with the
system, is located at $\sim 1700$ AU from the primary. We estimate
the expected gamma-ray luminosity from the colliding winds region,
the terminal shock of the wind, and the unstable zone at the base
of the wind, and conclude that, under very reasonable assumptions,
Cyg OB2 No. 5 can generate about a half of the gamma-ray flux
detected from the positionally coincident source 3EG J2033+4118.
We suggest, then, that other O stars belonging to the association,
also placed within the 95 \% probability EGRET location contour,
could contribute to the observed gamma-ray flux.


\end{abstract}

\section{Introduction}

High-energy emission from early-type stars has been often
discussed from a theoretical point of view (e.g. Cass\'e \& Paul
1980, V\"olk \& Forman 1982, White 1985, Chen \& White 1991, White
\& Chen 1992, Eichler \& Usov 1993). The observational evidence
for its existence, however, has remained rather elusive till now.

The recent publication of the Third EGRET (3EG) Catalog by Hartman
et al. (1999) provides new information that can be very useful for
the identification of possible gamma-ray emitting stellar systems.
There are 170 unidentified sources in the new catalog, and
approximately half of them are concentrated near the galactic
plane (e.g. Gehrels et al. 2000). Romero, Benaglia \& Torres
(1999) have studied the positional coincidence between 3EG sources
and different types of galactic objects such as supernova remnants
(SNRs), OB star associations (usually considered as tracers of
pulsar concentrations), and massive stars like Wolf-Rayet (WR) and
Of stars, which have very strong stellar winds. They have found
that 10 gamma-ray sources in the 3EG catalog have WR or Of stars
within their 95 \% confidence location contours. The probability
of this being a mere effect of chance is in the range
$10^{-2}-10^{-3}$. Although such probabilities are not as low as
those found for SNRs and OB associations, they are suggestive
enough as to encourage an {\sl a posteriori} analysis of some
interesting candidates. In particular, Romero et al. (1999)
mentioned the possibility that the system Cyg OB2 No. 5 could be
physically linked to the source 3EG J2033+4118, and hence that it
could be the first stellar system to be detected at gamma-rays,
involving other than WR stars. In this paper we discuss whether
Cyg OB2 No. 5 can actually generate the observed gamma-ray flux
and which are the relevant radiation mechanisms for this system at
high energies. 

Our results will be relevant not only for this
particular system but also for other interacting binaries like WR
140 (Romero et al. 1999, Zhekov \& Skinner 2000), which is also
positionally coincident with a gamma-ray source, and WR 147, for
which there are detailed radio observations (Contreras \&
Rodr\'{\i}guez 1999). The interest of stellar systems as potential
gamma-ray sources has been recently reinforced by the suggested
association of 3EG J1824-1514 with an X-ray binary by Paredes et
al. (2000).

We shall begin our study by reviewing the main characteristics of
Cyg OB2 No. 5.

\section{The stellar system Cyg OB2 No. 5}

\begin{table*}[]
\begin{center}
\begin{tabular}{l l c c c c c c c}
\multicolumn{9}{l} {{\bf Table 1}. Stellar parameters}\cr
 \hline &&&&&&&&\\
 Star & Sp.Class. & $\log{T_{\rm eff}}$ & $R_*$     &$\log(L/{\rm L}_{\odot})$ &
$B_*$ & $\dot{M}$ & $v_{\infty}$ & $V_{\rm rot}$ \\
      &           &       &(R$_{\odot})$  &                        &
(G)   &($10^{-5}$M$_{\odot}/{\rm yr}$) & (km/s)     & (km/s)  \\
\hline &&&&&&&&\\ Primary & O7 Ianfp$^{(\rm a)}$ &4.6$^{(\rm c)}$
&34$^{(\rm d)}$ & 5.984$^{(\rm c)}$   & 2     & $2.5^{(\rm f)}$ &
2200$^{(\rm d)}$ & 180$^{(\rm d)}$\\ &&&&&&&&\\ Secondary & B0
V$^{(\rm b)}$ & 4.5$^{(\rm c)}$ &8.3$^{(\rm c)}$ & 4.881$^{(\rm
c)}$    & 1     & $0.2^{(\rm f)}$ & 1000$^{(\rm b)}$ & 150$^{(\rm
e)}$\\ &&&&&&&&\\ \hline \multicolumn{9}{l} {(a): Walborn 1973;
(b): Contreras et al. 1997; (c) Vacca et al. 1996; (d): Bieging et
al. 1989;}\cr \multicolumn{9}{l} {(e): Uesugi \& Fukuda 1982; (f):
See text.}\cr
\end{tabular}
\end{center}
\end{table*}

Cyg OB2 No. 5 (also known as V729 Cyg, BD +40\degr4220) is an
evolved contact binary system with a period of 6.6 days (Hall
1974) located at a distance of $\sim 1.8$ kpc (e.g. Waldron et al.
1998). The primary star in the system is a very massive star
classified by Walborn (1973) as O7 Ianfp, although it should be
taken into account that a sizable fraction of the line emission
arises in the interaction region between the two stars and is not
intrinsic to the stars themselves. Most of the spectral features
that justify the ``nfp'' tag are actually emission lines that
form, to a large extent, in the wind interaction region of the
close binary.

Bohannan and Conti (1976) also classified the secondary star as an
Of star with similar luminosity, since each star seems to
contribute to the He II $\lambda 4686$ emission. This result has
been recently reconsidered by Rauw et al. (1999), who have
presented a detailed analysis of an extensive set of spectroscopic
observations finding a new orbital solution. These authors showed
that the observed properties can be explained if the secondary is
a transition Ofpe/WN9 star, with a much slower terminal wind
velocity than the primary. Consequently, at some distance from the
system, the wind properties are basically determined by the
primary, with a terminal velocity of $v_{\infty}\sim 2200$ km
s$^{-1}$ (Bieging et al. 1989).

Cyg OB2 No. 5 is well known as a rather powerful and variable
radio source (Abbott et al. 1981; Persi et al. 1985, 1990;
Miralles et al. 1994). The radio source seems to switch between
states of high and low emission with a period of about 7 years.
Miralles et al. (1994) detected, in addition to a strong source
coincident with the contact binary, a weak radio component,
possibly nonthermal, located at $\sim$ 1 arcsecond towards the NE
of the binary system. They proposed that the radio variability is
produced by relativistic particles accelerated in the wind. The
weak radio component, however, is too far from the binary system
as to be the radio counterpart of the secondary. In its
approximate position, a star of magnitude 13-14 has been
previously reported by Herbig (1967).

Recently, Contreras et al. (1997) have obtained accurate optical
(HIPPARCOS and Roque de los Muchachos) and radio (VLA)
measurements that clearly show that the weak radio source is
nonthermal and does not coincide with the optical position of the
star detected by Herbig (1967). This star, which seems to be a B0
V star, is a third component of the system, with a long period.
The fact that the weak radio source is located between the contact
binary and the B0 V star, closer to the latter, was interpreted by
Contreras et al. (1997) as the effect of the particle acceleration
in the region of colliding winds. In this region, electrons
locally accelerated radiate synchrotron emission, which is
detected at centimeter wavelengths as the weak radio component.

The system Cyg OB2 No. 5, then, seems to be formed by three stars:
an O7 Ia + Ofpe/WN9 contact binary and a B0 V star in a larger
orbit. When evaluating the gamma-ray production in Cyg OB2 No. 5,
we shall consider the contact binary, whose winds are dominated by
the O7 Ia star, as the primary object in the presumed gamma-ray
emitting system, and the B0 V star as the secondary.

In Table 1 we summarize the stellar parameters adopted in this
paper. The mass loss rate of the primary has been derived from
H$\alpha$, infrared and radio observations, giving values in the
range 1.9 to $2.6\times10^{-5}$ M$_{\odot} {\rm yr}^{-1}$ (e.g.
Leitherer et al. 1982, Persi et al. 1990). Contreras et al. (1997)
propose an extreme $\dot{M} = 5.5\times10^{-5}$ M$_{\odot} {\rm
yr}^{-1}$ that should be considered rather as an upper limit. In
our present work we adopt an average value of $2.5\times10^{-5}$
M$_{\odot} {\rm yr}^{-1}$. The mass loss rate for the secondary
results from that adopted for the primary and geometrical
considerations on the location of the colliding wind region (see
Section 4.1). The effective temperatures and the stellar
luminosities are taken from Vacca et al. (1996). There are not
direct measurements of the stellar magnetic fields, although they
are usually thought to be of the order of a few Gauss (e.g. Maheswaran \&
Cassinelli 1992). We
have adopted values of 2 and 1 G for the primary and the
secondary, respectively. A larger field would produce a turnover
in the synchrotron emission of the primary at a few GHz that is
not observed, whereas a smaller value would shift this turnover to
hundreds of GHz, resulting in an excess of non-thermal emission at
high frequencies. Our assumptions are in agreement with White \&
Chen (1992) and allow a fair description of the observed radio
spectrum for the primary in terms of White's (1985) model (see
Section 4.3).

\section{The gamma-ray source 3EG J2033+4118}

According to the 3EG catalog (Hartman et al. 1999), the best
estimated position of 3EG J2033+4118 is at
$(l,\;b)\approx(80.27\degr,\;0.73\degr)$, i.e at 0.27 degrees from
Cyg OB2 No. 5 (see Figure 1). 
The summed EGRET flux ($E>100$ MeV)
for cycles from 1 to 4 (i.e. for the time span 1991-1995) is
$(73.0\pm 6.7)\times10^{-8}$ ph cm$^{-2}$ s$^{-1}$. If we assume
that the source is at the distance of Cyg OB2 No. 5 (1.8 kpc), an
integrated gamma-ray luminosity of $\sim 2.4 \times 10^{35}$ erg
s$^{-1}$ is implied for the EGRET energy range (100 MeV -- 20
GeV). The photon spectral index is $\Gamma=1.96\pm 0.10$
($N(E)\propto E^{-\Gamma}$). For all calculations we assume
$\Gamma=2$.

In order to estimate the degree of variability of the gamma-ray
emission we have adopted the method used by Zhang et al. (2000),
Romero et al. (2000) and Torres et al. (2000). 
A variability index $I$ is defined such
that $I=\mu_{\rm s}/<\!\mu\!>_{\rm p}$, where $\mu_{\rm s}= 100\;
\sigma <\!F\!>^{-1}$ is the fluctuation index of the gamma-ray
source and $<\!\mu\!>_{\rm p}$ is the averaged fluctuation index
for all known gamma-ray pulsars, which are usually considered as a
non-variable population. $F$ is the flux and $\sigma$ its standard
deviation. A variable source, then, should satisfy that $I>1$. For
3EG J2033+4118 we obtain $I=1.3$, so, taken into account the
errors involved ($1\sigma=0.5$), the source cannot be classified
as clearly variable, at least over the timescale of the
observations ($\sim 4$ yr). This do not preclude that the
source could experience variability on longer timescales, for
instance with a period similar to what is observed at radio
frequencies in Cyg OB No. 5. (i.e., $\sim 7$ yr). We shall have to
wait until the next generation of gamma-ray satellites to get a
confident evaluation of the temporal behaviour of this source.

\begin{figure}[htbp]
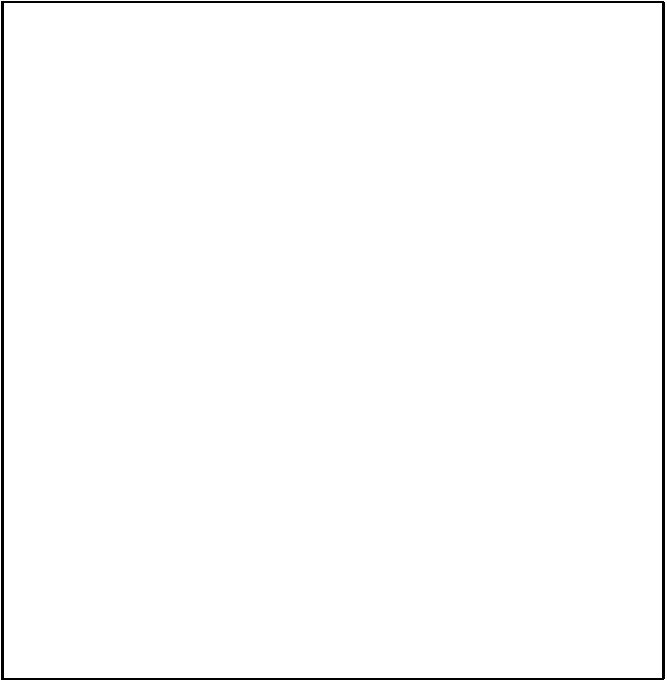

\picplace{9cm}
\caption{Location map of the gamma-ray source 3EG J2033+4118. The contours 
represent 50\%, 68\%, 95\% and 98\% statistical probability that the 
source lies within that contour. The position of the stellar system Cyg OB2
No. 5 is indicated with a star symbol}
\label{2033-2.eps}
\end{figure}

\section{Gamma-ray production in Cyg OB2 No. 5}

The main mechanisms that can generate gamma rays at energies
$E>100$ MeV in a stellar environment are inverse Compton (IC)
scattering, relativistic bremsstrahlung, and $\pi^0$-decay from
hadronic interactions. All these mechanisms require the existence
of a relativistic population of charged particles (electrons or
positrons in the first two cases, protons or ions in the latter).
Effective particle acceleration up to high energies requires, in
turn, the mediation of strong shock fronts (e.g. Bell 1978a,b;
Blandford and Ostriker 1978; Blandford 1980; Drury 1983): the
particles are scattered by magnetic irregularities carried by the
converging flow on either side of the shock wave, leading to a
first-order Fermi process that results in a power-law spectrum of
relativistic particles.

In a system such as Cyg OB2 No. 5 there are several places where
strong shocks can be formed. In particular, strong shocks are
expected in the colliding winds region located at some point
between the primary and secondary stars (i.e. between the contact
binary and the B0 V star), at the point where the momentum flux of
the primary (and stronger) wind decline to the value of
interstellar pressure (i.e. the terminal shock), and at the base
of the wind where line-driven instabilities are thought to occur.
These shocks locally re-accelerate the particles making them
energetic enough as to generate the gamma-rays. We shall discuss
below in major detail the possible gamma-ray production in each of
the mentioned regions.

\subsection{Region of colliding winds}

The supersonic winds from the primary and secondary stars flow
nearly radially out to the shocks. The geometry of the region
where the winds collide is described by Eichler \& Usov (1993) in
terms of the star separation $D$ and the parameter
$\eta=\dot{M_2}v_{\infty,2}/\dot{M_1}v_{\infty,1}$ as:

\be
r_1=\frac{1}{1+\eta^{1/2}}D,\;\;\;\;\;r_2=\frac{\eta^{1/2}}{1+\eta^{1/2}}D.
\ee Here, a subscript ``1'' stands for the primary whereas the
subscript ``2'' denotes the secondary. $\dot{M}$ is the mass loss
rate, $v_\infty$ is the wind terminal velocity, and $r$ the
distance from the star to the colliding winds region. Since
$\eta<1$, this region will be closer to the secondary.

Eichler \& Usov (1993) have shown that particles locally
accelerated at the strong shock front formed in the wind collision
can generate observable synchrotron radio emission. Contreras et
al. (1997) have identified the weak non-thermal radio source in
Cyg OB2 No. 5 with this emission. The same population of
relativistic electrons that emits the synchrotron radiation should
produce high-energy gamma-rays through IC interactions with the UV
photons from the secondary. The frequency of the IC photons is
given by $\nu_{\rm IC}=4/3\gamma^{2}\nu_{*}$, where $\gamma$ is
the Lorentz factor of the relativistic electrons ($\gamma \sim 3.5
\times 10^{3}$ for electrons that produce synchrotron photons of 5
GHz in a field of $\sim10^{-4}$ G) and $\nu_{*}$ is the frequency
of the seed stellar photons. The ratio of synchrotron to IC
luminosity is in this case (e.g. White \& Chen 1995):
\be
\frac{L_{\rm syn}}{L_{\rm IC}}=840\, \frac{B_{\rm s}^2
r_2^2}{L_2}, \label{LIC}\ee where $B_{\rm s}$ is the magnetic
field in Gauss, $r_2$ the distance to the secondary in AU, and
$L_2$ is its luminosity in L$_{\sun}$ units.

From geometric considerations, Contreras et al. (1997) derive
$\eta\sim0.04$, and then $r_2\sim300$ AU. The integrated
synchrotron luminosity between the two observing frequencies of
the nonthermal radio component is $L_{\rm
syn}\sim4.5\times10^{27}$ erg s$^{-1}$. Taken into account that
for a B0 V star the peak of the photon distribution is at
$\nu_*\sim2\times10^{15}$ Hz, we have that only electrons with
Lorentz factors in the range $3.0\times10^{3}-4.3\times10^{4}$
will contribute to the IC gamma-ray flux in the EGRET energy
range.

In order to estimate the total synchrotron luminosity of these
particles, we must first evaluate the magnetic field at the
colliding winds region. We shall assume that the external magnetic
field of the star in the absence of stellar wind is dipole, and
that in the presence of wind it obeys the standard $B\propto r^{-1}$ radial
dependence for large $r$ given by Eichler \& Usov (1993):

\be
B\approx B_{*} \frac{V_{\rm
rot}}{v_{\infty}}\frac{R_{*}^{2}}{r_{\rm A }\,r}\,,\;\;\;{\rm
valid\; for}\;\; r>R_{*}\frac{v_{\infty}}{V_{\rm rot}},
\label{B}\ee where $R_{*}$ is the stellar radius, $B_{*}$ the
surface magnetic field, $V_{\rm rot}$ the surface rotation
velocity, and $r_{\rm A}\approx R_{*}$ the Alfv\'en radius. Using
the parameters listed in Table 1 we find that in the colliding
winds region the field is $B\sim 2 \times 10^{-5}$ G, a value
about a factor 10 larger than the average magnetic field in the
standard ISM (Spitzer 1998). We shall adopt, taking into account
the compression produced by the strong shock formed in the wind
collision, a field $B_{\rm s} = 10^{-4}$ G. Then, the integrated
synchrotron luminosity of the electrons that contribute to the IC
gamma-ray luminosity is $L_{\rm syn}\approx5\times 10^{29}$ erg
s$^{-1}$. From Eq. (2), we obtain  that the corresponding inverse
Compton luminosity with seed photons from the secondary in the
EGRET energy range is $\sim5\times10^{34}$ erg s$^{-1}$. If we
also consider the contribution from the photon field of the
primary, we get a total IC luminosity of the colliding wind region
of $\sim 8\times10^{34}$ erg s$^{-1}$.

Additional contributions from bremsstrahlung and
 hadron
interactions can be expected from the colliding winds region. The
density of particles in this region is (Eichler \& Usov 1993):
\begin{eqnarray}
\label{n} n&\simeq&\frac{\dot{M}_{1}}{4\pi
r_{1}^2v_{\infty,1}m_{\rm p}\mu}\\
&\simeq&\frac{3\times10^{9}}{\mu}\left(\frac{\dot{ M}_{1}}{2\times
10^{-5}{\rm M}_{\sun}/{\rm yr} }
\right)\left(\frac{r_{1}}{10^{13}{\rm cm}}\right)^{-2}\nonumber\\
&& \times \left(\frac{v_{\infty,1}}{2\times10^{8}{\rm cm/s}
}\right)^{-1} {\rm cm}^{-3}. \nonumber
\end{eqnarray}
Here, $m_{\rm p}$ is the proton mass and $\mu$ is the mean
molecular weight of the gas, taken as 1.3. Using Eq. (4) and
assuming again a compression factor of 4 for the post-shock
region, we estimate a number density $n \simeq2.9\times10^{3}$
cm$^{-3}$. This means that the total mass in the colliding winds
region is $\sim2.2\times10^{-6}$ M$_{\sun}$. In this calculation
we have assumed a spherical geometry with a radius equal to the
average value of the semi-axes in Contreras et al.'s (1997) 6-cm
elliptical image of the secondary radio source.

The contribution from $\pi^{0}$-decays, consequently, should be
very low, even if an important enhancement in the protonic cosmic
ray energy density is occurring due to local reacceleration. We
estimate a luminosity of $\sim 5.2\times 10^{24}$ erg s$^{-1}$ for
a cosmic ray enhancement of a factor 10, using equation (19) in
Aharonian \& Atoyan (1996).

In the case of relativistic bremsstrahlung, which is produced in
the electrostatic fields of local ions by the same electronic
population that emits the observed synchrotron radiation, we have
that the ratio of gamma-ray bremsstrahlung ($F_{\gamma}$) to synchrotron radio
($S_{\rm syn}$) fluxes is given by (see Pollock 1985):

\begin{eqnarray}\label{bremm} R&=&\frac{F_{\gamma}(>100\;{ \rm MeV})}{S_{\rm
syn}}\\ &\simeq&1.4 \times 10^{-13}\left( \frac{n}{\rm
cm^{-3}}\right)\left(\frac{B_{\rm s}}{\mu \rm
G}\right)^{-3/2}\left(\frac{\nu}{\rm Hz}\right)^{1/2}\frac{\rm
cm^{-2}s^{-1}}{\rm Jy}. \nonumber
\end{eqnarray}

The above expression has been derived for a canonical power-law
electron spectrum of index $p=2$ ($N_{\rm e}(E) \propto E^{-p}$).
Using the observed radio flux at 5 GHz, and the same physical
parameters than in the case of the $\pi^{0}$-decays, we obtain a
bremsstrahlung contribution of $\sim 3.4 \times 10^{30}$ erg
s$^{-1}$. This value is more than a thousand times smaller than
the contribution from the inverse Compton emission, that clearly
dominates the high-energy radiation from this region.

We conclude that, according to the available radio observations, a
significant part of the observed gamma-ray flux from 3EG
J2033+4118 might be produced in the colliding winds region of Cyg
OB2 No. 5.

\subsection{Terminal shock}

Particle acceleration and gamma-ray production in the shocked
region at the boundary between a supersonic stellar wind and the
surrounding medium has been discussed in detail by Cass\'e \& Paul
(1980) and V\"olk \& Forman (1982). White (1985) showed that
energetic particles from the star can be injected into the
acceleration region despite the strong adiabatic losses. At the
terminal shock, the charged particles are accelerated to a power
law distribution with index $2$. Observable gamma-ray radiation
can be generated if there is a nearby source of photons to
interact with the electrons or nearby clouds that might be
``illuminated'' by the relativistic protons. In the case of Cyg
OB2 No. 5, if there is a contribution from the terminal shock to
the total gamma-ray luminosity, it should come from the latter
mechanism because there is no evidence of synchrotron emission
from the terminal shock that reveals a particularly high energy
density of relativistic electrons. On the contrary, there is
evidence of several compact clouds in the region. Dobashi et al.
(1994, 1996) have made an extensive study of molecular clouds in
the Cygnus region. They have found 159 molecular clouds through
large-scale $^{13}$CO observations in the region. The mass
spectrum of these clouds is found to be well approximated by a
power law $dN/dM_{\rm cl}\propto M_{\rm cl}^{-1.6}$, for $M_{\rm
cl}>100$ M$_{\sun}$. In the neighborhood of Cyg OB2 No. 5, four
clouds with estimated masses of $>230$, $>6750$, 1530, and $<240$
M$_{\sun}$ have been detected. The presence of many protostellar
candidates detected by IRAS also indicates a large number of
small, dense clouds. It is not unreasonable, then, to assume that
at least one small cloud of, say, 100 M$_{\sun}$ could be
illuminated by protons locally accelerated in the terminal shock
of the wind, at a few pc from the primary star. If the local
density of cosmic rays is enhanced in a factor 10 from the value
observed in the Earth vicinity (e.g. Dermer 1986), we can expect
an integrated gamma-ray luminosity of $\sim 2.3 \times 10^{32}$
erg s$^{-1}$ in the EGRET range, with the conservative assumption
that only a small cloud of 100 M$_{\sun}$ is overtaken.

\subsection{Base of the wind}

The winds of O and B stars are radiatively driven by absorption in
spectral lines and are prone to undergo instabilities that can
grow up to form strong shocks at the base of the outflow (e.g.
Lucy \& White 1980, Lucy 1982, Lamers \& Casinelli 1999). These
shocks can accelerate particles to relativistic energies. If the
 particles are electrons, syncrhotron and IC radiation can be
produced. White (1985) have calculated the electron energy
spectrum and the associated synchrotron spectrum, and have shown
that these spectra can be quite different from the naive pure
power-law case. In particular, the wind synchrotron spectrum rises
with frequency as $S_{\nu}\propto\nu^{1.9 \beta/6}$, with
$\beta=1.6$ in the strong shock limit, for frequencies below a
turnover $\nu_{\rm t}$, and is nearly flat ($S_{\nu}\propto \ln
\nu_{\rm t}$) from the turnover to the high energy cutoff. This
theoretical synchrotron spectrum agrees notably well with the
radio spectrum of the primary in Cyg OB2 No. 5, which presents a
spectrum $\propto\nu^{0.54}$ between 4.8 and 8.4 GHz (Waldron et
al. 1998), that flattens at higher frequencies (Contreras et al.
1996). The value of $\nu_{\rm t}$, given by equation (45) of White
(1985), is highly sensitive to small changes in the stellar
parameters. For the values listed in Table 1 for the primary in
Cyg OB2 No. 5 we obtain $\nu_{\rm t}\approx10$ GHz, in good
agreement with the radio data.

The high-energy cutoff for charged particles accelerated in the
vicinity of a star is (White 1985):

\be
E_{\rm max}\approx 1\; {\rm GeV}\left(\frac{v_{\infty,1}}{2500\;
{\rm km/s}}\right)^{-1/2}\left(\frac{B_{*}}{{\rm G}}\right)
\left(\frac{R_{*}}{35 {\rm R}_{\odot}}\right), \label{Emax} \ee
which yields $E_{\rm max}\approx 2$ GeV for the primary in Cyg OB2
No. 5 (i.e. Lorentz factors $\gamma\sim4\times10^{3}$). This upper
limit, imposed by geometric factors, is valid for both electrons
and ions. However, in the case of electrons, radiative losses
mainly due to IC scattering impose additional constraints (Chen \&
White 1991), and hence the effective high-energy cutoff should
occur at $\gamma\la10^{3}$ for them. Consequently,
electron-induced gamma-rays in the base of the wind have energies
$\la100$ MeV and are below EGRET energy threshold, as noticed by
Chen \& White (1991).

\begin{table*}[htbp]
\begin{center}
\begin{tabular}{ l | l | c | c }
\multicolumn{4}{l} {{\bf Table 2}. Gamma-ray production in Cygnus
OB2 No. 5 for EGRET energy range}\cr \multicolumn{4}{l} {(100 MeV
$<$ $E$ $<$  20 GeV)}\cr \hline &&&\\
 Region    &  Mechanism      & Expected luminosity & Observed  luminosity \\
           &                 &  (erg/s)   &    (erg/s)   \\
&&&\\
\hline
&&&\\
           & Inverse Compton scattering & $\sim8\times 10^{34}$& \\
&&&\\ \cline{2-3} &&&\\ Winds collision     & Relativistic
bremsstrahlung  &  $\sim3.4\times 10^{30}$& \\ &&&\\ \cline{2-3}
&&&\\
           & Neutral pion decay &   $\sim5.2\times 10^{24}$& \\
&&& $\sim2.4\times 10^{35}$ \\ \cline{1-3} &&&\\ Terminal shock &
Neutral pion decay   &    $\sim2.3\times 10^{32}$& \\ &&&\\
\cline{1-3} &&&\\
           & Inverse Compton scattering & -- &\\
Base of the wind &&&\\ \cline{2-3} &&&\\
          &  Neutral pion decay$^1$   &    $\sim5\times 10^{34}$& \\
&&&\\ \hline \multicolumn{4}{l} {1: White \& Chen 1992}\cr
\end{tabular}
\end{center}
\end{table*}

Relativistic ions, on the contrary, can produce substancial
gamma-ray luminosities through hadron interactions with the
thermal wind nuclei and the subsequent $\pi^{0}$-decays (White \&
Chen 1992). The resulting high-energy emission has a peak at $\sim
70$ MeV and then can significantly contribute to the gamma-ray
flux detected by EGRET.

The calculation of the $\pi^{0}$ gamma-ray luminosity of the
primary star in Cyg OB2 No. 5 has already been carried out by
White \& Chen (1992), using the empirical formula for the
$\pi^{0}$ production spectrum given by Stephens \& Badhwar (1981).
The expected luminosity is $\sim 5 \times 10^{34}$ erg s$^{-1}$,
similar to the IC luminosity of the colliding winds region.

\subsection{Summary of results}

In Table 2 we present a summary of the different contributions to
the total gamma-ray luminosity of Cyg OB2 No. 5 in the EGRET
energy range. The main contributions are IC emission from the
colliding winds region and $\pi^{0}$-decay emission from the inner
region of the primary stellar wind. These mechanisms can provide
about a half of the inferred luminosity of the source 3EG
J2033+4118 (i.e. $\sim1.3\times10^{35}$ erg s$^{-1}$). The
observed gamma-ray flux could be explained if either (1) the mass
illuminated by relativistic protons accelerated at the terminal
shock of the wind is a factor $\sim50$ larger than what we have
assumed (i.e. if it is $\sim$5000 M$_{\odot}$), (2) there are
other gamma-ray sources within the EGRET 95 \% probability
location contour or (3) there is an important contribution from
the colliding winds region within the primary (i.e. in the close
binary). This latter possibility cannot be quantitatively
discussed because of the absence of observational constraints on
the efficiency of the particle acceleration at the unresolved
interacting region. The second possiblity, namely that the EGRET
detection could actually be a composed source, will be briefly
discussed below.

\subsection{Other potential gamma-ray sources in the field}

There are, in addition to Cyg OB2 No. 5, other massive stars of
the same association within the positional error box of  3EG
J2033+4118, like stars No. 7, 8A, 9, and 11. These stars have
similar parameters to the primary in Cyg OB2 No. 5 and are
potential sources of $\pi^{0}$-decay gamma-rays. In particular,
Cyg OB2 No.8A and No.9 are non-thermal radio emitters (Contreras
et al. 1996, Waldron et al. 1998). White \& Chen (1992) have
estimated the individual gamma-ray luminosities at $E>100$ MeV of
these stars in $\sim 5\times 10^{34}$  erg s$^{-1}$. The combined
effect of these stars, plus the emission from the colliding winds
region in Cyg OB2 No. 5, can easily explain the flux detected by
EGRET. The absence of any other strong radio source in the field
also supports this conjecture, which was advanced by White \&
Cheng (1992) prior to the detection of non-thermal radio emission
from Cyg OB2 No. 5 (see also Chen et al. 1996).

\section{Further comments}

\subsection{High-energy tail of the gamma-ray emision}

In the colliding winds region of Cyg OB2 No. 5 particles are
accelerated at the shock front up to a high-energy cutoff that can
be estimated assuming that the diffusion length is equal to the
gyral radius. This yields,

\be
E_{\rm max}=1.3\times 10^{3}\;{\rm GeV}\;\left(\frac{v_{\rm
s}B}{L_{2}}\right)^{1/2}r_{2}, \ee where $v_{\rm s}$ is the shock
velocity in km s$ ^{-1}$. If $v_{\rm s} = v_{\infty,1}$ and $B =
B_{\rm s}$, we obtain $E_{\rm max}\sim0.7$ TeV. The IC high-energy
emission, then, extends up to energies $\sim 20$ TeV. The expected
gamma-ray luminosity at $E>1$ TeV is $5 \times 10^{30}$ erg
s$^{-1}$, which, at a distance of 1.8 kpc gives a photon flux of
$1.4 \times 10^{-11}$ cm$^{-2}$s$^{-1}$. This flux could be
detected by planned gamma-ray ground arrays like VERITAS (Weekes
et al. 2000).

\subsection{A few comments on the soft gamma-ray emission}

The colliding winds region can be an important source of soft
gamma-ray photons. The IC flux density around 10 MeV is $8\times
10^{-5}$ ph  cm$^{-2}$ s$^{-1}$ and could be detected, in
principle, by the Imager on Board INTEGRAL Satellite (IBIS), to be
launched by the European Space Agency in April 2002 (Winkler \&
Hermsen 2000). IBIS is a coded mask telescope designed to produce
sky images in the 20 KeV -- 10 MeV band with an angular resolution
of 12 arcminutes. It will also have a good spectral capability.
The point source location accuracy for a relatively strong source
will be around 1 arcminute. Consequently, IBIS will provide
excellent diagnosis to test the models for gamma-ray production in
Cyg OB2 No. 5 discussed in this paper. In particular, it will
allow to measure the soft gamma-ray spectrum of the colliding
winds and differentiate the emission of Cyg OB2 No. 5 from that
produced in the neighboring stars, which are located at an angular
distance $\geq 9$ arcminutes. The main problem at these energies
is the subtraction of the background galactic emission;
notwithstanding, sufficiently accurate models of the diffuse
radiation fields at low latitudes will be soon available (e.g.
Bloemen et al. 2000). IC emission from the single stars should
also be important at MeV energies. The hadronic gamma-ray
component, however, should be observed with future higher energy
telescopes, like the planned GLAST satellite, to be launched by
NASA towards 2005.

\section{Conclusions}

We have presented an analysis of the possible production of
gamma-ray emission in the stellar system Cyg OB2 No. 5 taken into
account the constraints imposed by recent high-resolution radio
observations. We have estimated the gamma-ray luminosity generated
by different types of mechanisms and compared them with the
luminosity inferred from EGRET observations. We have found that
about a half of the flux of the gamma-ray source 3EG J2033+4118
could arise from the colliding winds region (through IC
interactions)
 and the base of the wind (through
proton-proton interactions). It is possible, also, that the
remaining flux comes from other nearby early-type stars of the
same OB association within the positional error box of the EGRET
detection. We expect that forthcoming gamma-ray missions with good
imaging capabilities, like the imminent ESA's INTEGRAL spacecraft,
can clarify this point.

\begin{acknowledgements}

 This work has been supported by the Argentine agencies CONICET (PIP 0430/98)
 and ANPCT (PICT 98 No. 03-04881), as well as by Fundaci\'on
 Antorchas.We thank Gerry Skinner and Gregor Rauw for valuable
 comments.

\end{acknowledgements}


{}

\end{document}